\documentclass[showpacs,twocolumn,prb]{revtex4}
\usepackage{bm}
\usepackage{graphicx}
\usepackage{amsmath}
\usepackage{amssymb}
\usepackage{color}

\newcommand{\bra}[1]{\left\langle {#1} \right\vert}
\newcommand{\ket}[1]{\left\vert {#1} \right\rangle}

\begin{document}

\title{Electric field effect on electron spin
splitting in SiGe/Si quantum wells}

\author{M.O.~Nestoklon} 
\author{E.L.~Ivchenko}
\affiliation{A.F.~Ioffe Physico-Technical Institute, Russian
Academy of Sciences, St. Petersburg 194021, Russia}

\author{J.-M.~Jancu}
\author{P.~Voisin}

\affiliation{Laboratoire de Nanophotonique et Nanostructures, 91460
Marcoussis, France}

\begin{abstract}
{Effect of electric field on spin splitting in SiGe quantum wells
(QWs) has been studied theoretically. Microscopical calculations 
of valley and spin splittings are performed in the efficient
$sp^3d^5s^*$ tight-binding model. In accordance with the symmetry
considerations, the electric field not only modifies the interface
induced spin splitting but gives rise to a Rashba-like
contribution to the effective two-dimensional electron
Hamiltonian. Both the valley and spin splittings oscillate as a
function of the QW width due to inter-valley reflection of the
electron wave off the interfaces. The oscillations of splitting
are suppressed in rather low electric fields. The tight-binding
calculations have been analyzed by using the generalized envelope
function approximation extended to asymmetrical QWs.}
\end{abstract}
\maketitle

\section{Introduction}
Understanding the details of semiconductor hetero\-structure electronic 
properties has been a key at each stage of their applications in the 
field of information and communication technologies. Presently, there 
is a broad interest in spin-dependent properties because they have a 
potential for novel ``spintronic" devices, and beyond, because they 
govern for a large part the possible development of semiconductor-based 
quantum information processing. Spin splitting of electron dispersion 
relations arises from the combination of spin-orbit coupling and 
inversion asymmetry. Besides the contributions of bulk inversion 
asymmetry (BIA) first discussed by Dresselhaus in 1954 and the 
structure inversion asymmetry (SIA) introduced by Rashba,\cite{Rashba,Bychkov}
 the existence of a contribution due to the breakdown of roto-inversion 
symmetry at an interface between two semiconductors was first
suggested by Vervoort et al. in the late
90s.\cite{Vervoort1,Vervoort2} This ``interface inversion
asymmetry'' (IIA) term was further documented both by
group-theoretical analysis,\cite{Ivchenko_Kaminski}
envelope-function calculations\cite{Vervoort2} and measurements of 
circular polarization relaxation in quantum wells (QWs) based on
various III-V semiconductors.\cite{Guettler,Olesberg,Hall}
However, in these cases, the interface contribution appears in
combination with BIA and SIA. Pure IIA can exist alone or together
with SIA in heterostructures of centro-symmetric semiconductors
like Si-Ge QWs. Previous works\cite{Golub} have established the
symmetry properties specific to this system where electrons lie in
states originating from the bulk X$_{\rm{z}}$ valleys. A general
feature of zone-edge conduction valleys in bulk materials is their
degeneracy (6 for X-valleys, 4 for L-valleys) that gives rise to
strong valley-coupling when they are folded onto the
two-dimensional Brillouin zone of a
QW.\cite{Boykin_vo,Our_SiGe,Jancu_GaSb,Valavanis,Virgilio,Friesen}
Valley coupling is another manifestation of  the local,
three-dimensional variations of crystal potential at semiconductor
interfaces and quantitative estimates require atomistic
information which is not available within the $\bm{k}\cdot\bm{p}$
theoretical framework. Parameters describing valley coupling must
be extracted from microscopic approaches such as ab-initio
calculations or modelizations using empirical parametrizations
like atomistic pseudopotentials or tight-binding approach. The
valley coupling strongly depends on the crystalline growth
direction and shows an oscillating behaviour as a function of the
number of monolayers forming the X$_{\rm{z}}$-valley or L-valley
quantum well. It also depends on the overall symmetry of the
quantum well and for this reason, it can be modified by an
external electric field. The spin-splittings of in-plane
dispersion relations in such systems results from the interplay of
valley coupling and spin-dependent terms in the electron
Hamiltonian. The case of L-valley QWs formed in the GaSb-AlSb
system and grown along the [001] direction was first discussed in
some details by Jancu et al.\cite{Jancu_GaSb} In that case, the
leading terms come from BIA invariants specific to the L valleys
in combination with the L-valley coupling. For Si-Ge (001)-grown
QWs the interplay of IIA with X$_{\rm{z}}$ valley coupling was
examined from the $\bm{k}\cdot\bm{p}$ theory point of view and
semi-quantitative estimates were discussed in the frame of
tight-binding calculations based on the $sp^3s^*$
model.\cite{Our_SiGe} However, it is well known that this simple
model cannot reproduce quantitatively the properties of zone edge
valleys such as effective masses and dipole matrix element. This
difficulty was solved in the late nineties by the introduction of
the extended basis $sp^3d^5s^*$ tight binding model.\cite{Jancu}
More recently, progress in the parametrization of this model have
led to essentially perfect description of the electronic
properties of bulk Ge\cite{Jancu_strain} and Si.\cite{Sacconi} In
this work, we use the advanced tight-binding model in combination
with the envelope function approach and calculate the conduction
band spin splitting resulting from the interplay of valley
coupling with the IIA and electric-field effects in Si/SiGe
quantum wells.

\section{Point-group symmetry analysis}
In the virtual-crystal approximation for SiGe alloys, an ideal
(001)-grown SiGe/Si/SiGe QW structure with an odd number $N$ of Si
atomic planes has the point-group symmetry D$_{2d}$ which allows
the spin-dependent term $\alpha (\sigma_x k_x - \sigma_y k_y)$ in
the electron effective Hamiltonian, where $\sigma_x, \sigma_y$ are
the Pauli spin matrices, ${\bm k}$ is the two-dim\-en\-sion\-al
wave vector with the in-plane components $k_x, k_y$, and $x
\parallel [100], y \parallel [010]$. The QW structures with even
$N$ have the D$_{2h}$ point symmetry containing the
space-inversion center, the constant $\alpha$ is zero and the
two-dimensional electronic states are doubly degenerate. Under an
electric field ${\bm F} = (0,0,F_z)$ applied along the growth
direction $z$ the symmetry of QW structures with both odd and even
numbers of Si monoatomic layers reduces to the C$_{2v}$ point
group and the spin-dependent linear-$\bm{k}$ Hamiltonian becomes
\begin{equation} \label{hamso}
\mathcal{H}^{(1)}({\bm k}) = \alpha (\sigma_x k_x - \sigma_y k_y)
+ \beta (\sigma_x k_y - \sigma_y k_x)\:,
\end{equation}
where the second contribution is usually called the Rashba term
(or the SIA term). In order to establish the parity of $\alpha,
\beta$ with respect to inversion of the electric field we note
that, in the D$_{2d}$ group, the combination $h({\bm k}) =
\sigma_x k_x - \sigma_y k_y$ as well as even powers of $F_z$ are
invariants while both the combination $h'({\bm k}) = \sigma_x k_y
- \sigma_y k_x$ and odd powers of $F_z$ transform according to the
same representation B$_2$ (in notations of Ref.~\onlinecite{Bir}).
Therefore, for structures with odd $N$, the coefficients $\alpha$
and $\beta$ are, respectively, even and odd functions of $F_z$.
They can be presented as
\begin{eqnarray} \label{evenodd}
\alpha(F_z; {\rm odd}~ N) &=& \alpha_0 + c_{\alpha}^{(2)} F^2_z +
c_{\alpha}^{(4)} F^4_z +...\:,\\ \beta(F_z; {\rm odd}~ N) &=&
c_{\beta}^{(1)} F_z + c_{\beta}^{(3)} F^3_z +...\:, \nonumber
\end{eqnarray}
where $\alpha_0 \equiv \alpha(0)$ and $c_{\alpha}^{(2n)},
c_{\beta}^{(2n+1)}$ are field-independent coefficients. Similarly,
for structures with even $N$, the linear-in-${\bm k}$
spin-dependent Hamiltonian can be presented in the form
\begin{equation}\label{D2h_field_Hso}
\mathcal{H}^{(1)}(F_z; {\rm even}~ N) = 
F_z\ [ C_1 h({\bm k}) + C_2 h'({\bm k}) ]\ ,
\end{equation}
where $C_1, C_2$ are even functions of $F_z$. The above
representation follows immediately if we take into account that,
with respect to operations of the D$_{2h}$ group, both $h({\bm
k})$ and $h'({\bm k})$ transform in the same way as the component
$F_z$ does.

The aim of this work is to calculate and analyze the electric
field dependencies of $\alpha$ and $\beta$. For this purpose we
use the precise nearest-neighbor $sp^3d^5s^*$ tight-binding
model~\cite{Jancu} and calculate valley and spin splittings in
symmetrical QWs in the absence and presence of an external
electric field.

\section{Tight-binding model}
To calculate electron subband splittings, we use the $sp^3d^5s^*$
tight-binding theory elaborated by Jancu et al.\cite{Jancu} It
perfectly reproduces band structure of indirect bulk
semiconductors as well as electron effective masses, etc. In
particular the parametrization used in this work reproduces the
value $k_0$=85\% of the conduction band minimum in Si, which was
considered as a challenge.\cite{Klimeck_Si} One of the main
advantages of this method is a very straightforward treatment of
nanostructures.

In Ref.~\onlinecite{Our_SiGe}, we estimated the electron spin
splitting in symmetrical SiGe QWs using a less detailed
tight-binding model, namely, the $sp^3s^*$ model, which allowed us
to understand the main qualitative features of spin splitting as
well as to demonstrate the possible observability of this effect
in Si/SiGe heterostructures.

In the tight-binding model the electron wave function is written
as a linear combination of atomic orbitals~\cite{SlaterKoster}
\begin{equation}\label{TB_wavefunction}
\ket{\psi,\bm{r}}=\sum_{n,\nu}C_{n,\nu}\ket{\Psi_{\nu},\bm{r}-\bm{r}_n},
\end{equation}
where $n$ enumerates atoms in the structure, $\nu$ runs through
the set of spinor orbitals at the $n$th atom. In the $sp^3d^5s^*$
model, this set includes the orbitals $s$, $p_{\eta}$ $( \eta =
x,y,z)$, $d_{\xi}$ $(\xi = yz, xz, yz, x^2-y^2 ,2z^2-x^2-y^2)$ and
$s^*$ multiplied by the spinors $\uparrow$ and $\downarrow$. We
assume the basic orbital functions to be orthogonal. Anyway, this
can be achieved using the L{\"{o}}wdin orthogonalization
procedure. \cite{Lowdin} Thus, the tight-binding Hamiltonian is
presented as a multicomponent matrix and the Schr{\"{o}}dinger
equation as an eigenvalue problem
\begin{equation}\label{TB_hamiltonian}
\sum\limits_{n', \nu'} \bra{\Psi_{n, \nu}} H \ket{\Psi_{n', \nu'}}
C_{n', \nu'} = E C_{n, \nu}\ ,
\end{equation}
where $\ket{\Psi_{n,\nu}} = \ket{\Psi_{\nu},\bm{r}-\bm{r}_n}$. The
Hamiltonian matrix elements depend on the relative position of
atoms, $\bm{r}_n -\bm{r}_{n'}$, and chemical type of atoms $n$ and
$n'$. We use here the nearest neighbour approximation where the
matrix elements differ from zero only for neighbouring atoms. The
detailed procedure of constructing the tight-binding Hamiltonian
can be found in Ref.~\onlinecite{SlaterKoster}. Strain effects can
be included by scaling the matrix elements with respect to the
bond-angle distortions and bond-length changes. \cite{TB_strain}

For the SiGe alloy we use the virtual crystal approximation (VCA)
in order to concentrate on the intrinsic structure symmetry thus
neglecting all effects of disorder. Tight-binding parameters were 
optimized to carefully reproduce alloy band structure. We treat
strain in two independent ways: first, atomic positions used in
calculations are chosen by using Van de Walle's
model.\cite{VandeWalle} We have also applied Keating's Valence
Force Field (VFF) model \cite{Keating} with the SiGe parameters
from Ref.~\onlinecite{VFF_Zunger} and found no difference between
the continuous and atomistic approaches.

In addition to the strain dependence of tight-binding parameters
we have corrected the structure potential (see below) with respect
to experimentally observed Si/Si$_{1-x}$Ge$_{x}$ conduction-band
offset.\cite{Rieger,Schaffler}

We treat an electric field in the tight-binding approach in the
following way:\cite{Boykin_Vogl} the diagonal matrix elements of
the tight-binding Hamiltonian are shifted due to the potential of
the applied electric field
\begin{multline}\label{TB_field}
\bra{\Psi_{n, \nu}} H \ket{\Psi_{n', \nu'}} \\= \bra{\Psi_{n, \nu}}
H \ket{\Psi_{n', \nu'}}_{U=0} + U(\bm{r}_n)\ \delta_{nn'}\
\delta_{\nu \nu'}\ ,
\end{multline}
where $U(\bm{r})$ is the electric potential energy.

Since we are interested in the in-plane dispersion of free
electrons in a heterostructure we impose periodical boundary
conditions in the interface plane (001). Because of the
periodicity in the [100] and [010] directions we can introduce the
in-plane wave vector $\bm{k}$ and, for a given value of $\bm{k}$,
construct the tight-binding Hamiltonian with a discrete spectrum.
For the sake of numerical simplicity we also use periodic boundary
conditions along the growth direction [001] taking the barrier
layers thick enough to exclude the influence of their thickness on
the calculated values of $\alpha$ and $\beta$. 
It follows immediately from the band structure of silicon and
SiGe/Si/SiGe structure potential that, neglecting the valley
splitting, the electronic states with $k_x = k_y = 0$ are
four-fold degenerate. We will focus on the dispersion of the
lowest conduction subband $e1$. The interface-induced valley
mixing leads to a splitting of the state $\ket{e1, {\bm k}=0}$
into two spin-degenerate states denoted $+$ (upper subband) and
$-$ (lower subband). At nonzero ${\bm k}$ each subband, $+$ and
$-$, undergoes the spin-orbit splitting described by
Eq.~(\ref{hamso}) with the coefficients $\alpha_{\pm}$ and
$\beta_{\pm}$ for the valley-orbit split subbands $\pm$. It is
instructive to rewrite Eq.~(\ref{hamso}) in the coordinate frame
$x' \parallel [1\bar{1}0], y' \parallel [110]$ as follows
\begin{equation} \label{hamsoa}
\mathcal{H}^{(1)}({\bm k}) = (\alpha_{\pm} + \beta_{\pm})
\sigma_{x'} k_{y'} + (\alpha_{\pm} - \beta_{\pm}) \sigma_{y'}
k_{x'}\:.
\end{equation}
Let us introduce the energy difference $\Delta^{(\pm)}_{\rm so}
(\bm{k} \parallel [1\bar10])$ for the states $\ket{\pm, {\bm k}
\parallel [1\bar{1}0]}$ with the spin polarized parallel and antiparallel
to [110], and $\Delta^{(\pm)}_{\rm so} (\bm{k} \parallel [110])$
for the states $\ket{\pm, {\bm k} \parallel [110]}$ with the spin
polarized parallel and antiparallel to $[1\bar{1}0]$. The modulus
of $\Delta^{(\pm)}_{\rm so}(\bm{k})$ gives the spin splitting of
the $\pm$ subbands and the sign of $\Delta^{(\pm)}_{\rm
so}(\bm{k})$ determines the relative position of the split spin
sublevels. It follows from Eq.~(\ref{hamsoa}) and the definition
of $\Delta^{(\pm)}_{\rm so} (\bm{k})$ that the constants
$\alpha_{\pm}, \beta_{\pm}$ can be found from
\begin{eqnarray} \label{ab}
\alpha_{\pm} = \lim_{k\rightarrow + 0} \frac{ \Delta^{(\pm)}_{\rm
so} (\bm{k} \parallel [110] ) + \Delta^{(\pm)}_{\rm
so} (\bm{k} \parallel [1\bar10]) }{ 4|\bm{k}|}\: ,\\
\beta_{\pm} = \lim_{k\rightarrow + 0} \frac{\Delta^{(\pm)}_{\rm
so} (\bm{k} \parallel [110]) - \Delta^{(\pm)}_{\rm so}
(\bm{k}\parallel [1\bar10])}{4|\bm{k}|}\:. \nonumber
\end{eqnarray}

Also it should be pointed out that, since the studied QWs are
quite shallow, the electron dispersion should be treated with
care. To avoid non-linear effects, very small values of $k_x, k_y$
should be considered.

\section{Results and discussion}
\subsection{Numerical $sp^3d^5s^*$ model calculations: unbiased structure}

In order to test and improve our previous results we calculated
valley and spin splitting in symmetrical Si QW with 
Si$_{0.75}$Ge$_{0.25}$ barriers as a function of the QW width. For
SiGe composition we used optimized tight-binding parameters
precisely reproducing realistic alloy band structure. The strategy
for parameterization of the Ge-Si alloy in a virtual crystal
approximation is as follow: the parameters of Ge and Si
hydrostatically strained to the alloy parameter are first
calculated and linearly interpolated. The small remaining
differences with measured values are then corrected by fine tuning
of a few two-center parameters. For interface atoms we use linear
combination of pure Si parameters and alloy.

Tight-binding parameters are optimised for bulk materials. However, 
band offsets at the interfaces in the heterostructure
are also important. For conduction band offset we use Sh\"affler's paper~\cite{Schaffler}
as a reference. Thus, we take a value of 150~meV as a conduction band offset
for X$_{\rm{z}}$ valley electrons.

Figure 1 shows the zero-field results of calculation of (a) the
valley splitting $\Delta_{\rm v}$ and (b) the constants
$\alpha_{\pm}$ for the valley split $e1$ subbands in a symmetrical
single QW structure with $N$ Si atomic planes sandwiched between
the thick Si$_{0.75}$Ge$_{0.25}$ barriers. The splittings as a
function of $N$ exhibits oscillations, in agreement with
Refs.~\onlinecite{Our_SiGe,Boykin_vo,Valavanis,Friesen}. In Fig.~1
X-shaped crosses depicted as vortices of the broken line represent
the calculation in the envelope-function approximation, see below.
The broken line is drawn for the eye. Note that, in order to
simplify comparison with results obtained by other authors,
Fig.~1a illustrates the valley splitting not only for the QW width
region 15$\div$50 {\AA}~but also for the region 60$\div$70~{\AA}.

\begin{figure}
\includegraphics[width=.29\textwidth]{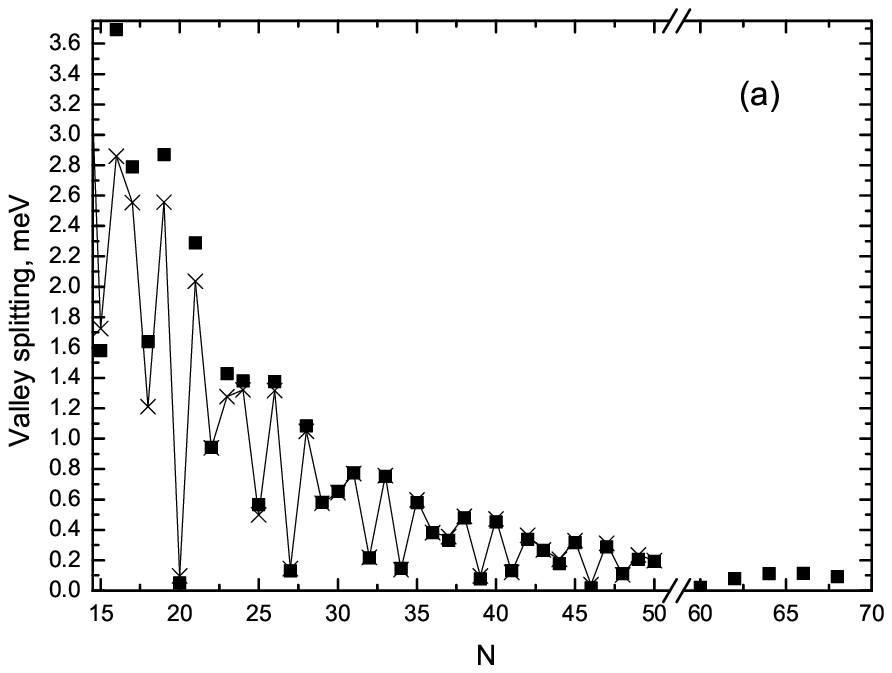}
\hspace{0.5cm}
\includegraphics[width=.33\textwidth]{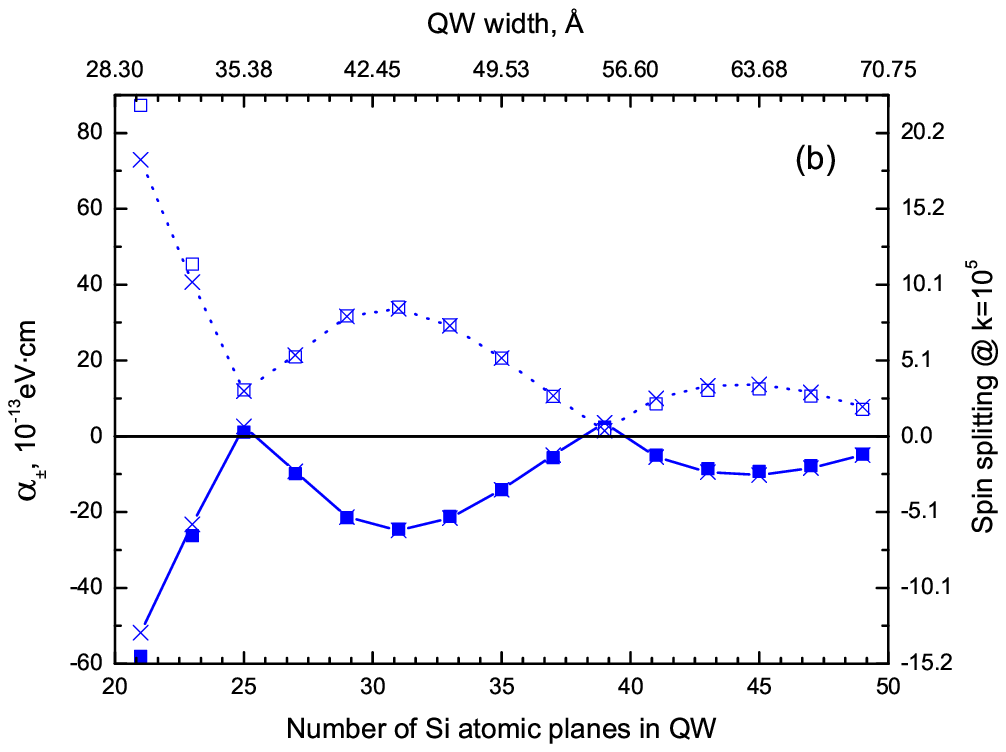}
\caption{(a) Valley splitting $\Delta_{\rm
v}$ as a function of $N$ in a
Si$_{1-x}$Ge$_x$/Si/Si$_{1-x}$Ge$_x$ ($x$ = 0.25) 
QW structure in the absence of an electric field. 
Solid squares and vortices of the broken line represent results of calculations using the tight-binding
method and envelope function approximation, respectively.
(b) Spin-splitting constants $\alpha_{\pm}$ versus the number 
$N$ of Si monoatomic layers in the same system (odd $N$ are taken in account
only). Spin splitting is shown by solid and open squares 
($sp^3d^5s^*$ tight-binding calculation, $\alpha_-$ and $\alpha_+$ respectively) 
and x-shaped crosses (envelope function approximation).}
\end{figure}

Results obtained in the framework of the advanced $sp^3d^5s^*$
tight-binding model show considerable difference with our previous
estimations.\cite{Our_SiGe} The valley splitting is significantly
smaller, its value decreases by a factor of 3, whereas the spin
splitting increases almost six times. This difference is not
unexpected since a careful tight-binding treatment of Si and its
compounds is possible in the $sp^3d^5s^*$ model only. In this
regard the goal of the previous paper \cite{Our_SiGe} was to
demonstrate that the effect of spin splitting in macroscopically
symmetrical Si quantum wells is measurable and to reveal the main
qualitative properties of this splitting.

The previous theoretical values for valley splitting were obtained
by both tight-binding\cite{Boykin_vo,Friesen} and pseudopotential
methods.\cite{Valavanis} Although the first two papers utilize the
method of calculation similar to that applied here, a
straightforward comparison is not possible due to different 
parametrizations of the Si$_{1-x}$Ge$_x$ alloy, different alloy
compositions ($x$ = 0.2 in Ref.~\onlinecite{Boykin_vo} and 0.3 in
Ref.~\onlinecite{Friesen}) and conduction band offsets used.
However, our results are in good agreement with the both
estimations. For example, for a QW containing 64 Si-atomic layers
(32 monomolecular layers, 9 nm) we obtain for the valley splitting
$\sim 0.11$ meV , while Refs.~\onlinecite{Boykin_vo} and
\onlinecite{Friesen} present the coinciding values of $\sim 0.2$.
Our analysis shows that the valley splitting is quite sensitive to
the SiGe alloy parameters. By using the linear combination of Si
and Ge tight-binding parameters for the alloy we could reproduce
values of the valley splitting obtained by Boykin et
al.\cite{Boykin_vo}

Comparison with Ref.~\onlinecite{Valavanis} is more
straightforward. Figure~1 in the cited paper shows dependence of
the valley splitting on the barrier Ge content for a 16 Si-atomic
layer QW calculated by the empirical pseudopotential method. In
particular, for the
Si$_{0.75}$Ge$_{0.25}$/Si/Si$_{0.75}$Ge$_{0.25}$ QW the valley
splitting of about 2.5 meV was obtained\cite{Valavanis} while our
estimation is 3.7 meV. This is a good agreement taking into
account that the two values are obtained in two completely
different approaches for quite narrow QWs where interface effects
are extremely important.

\begin{figure}
\includegraphics[width=.28\textwidth]{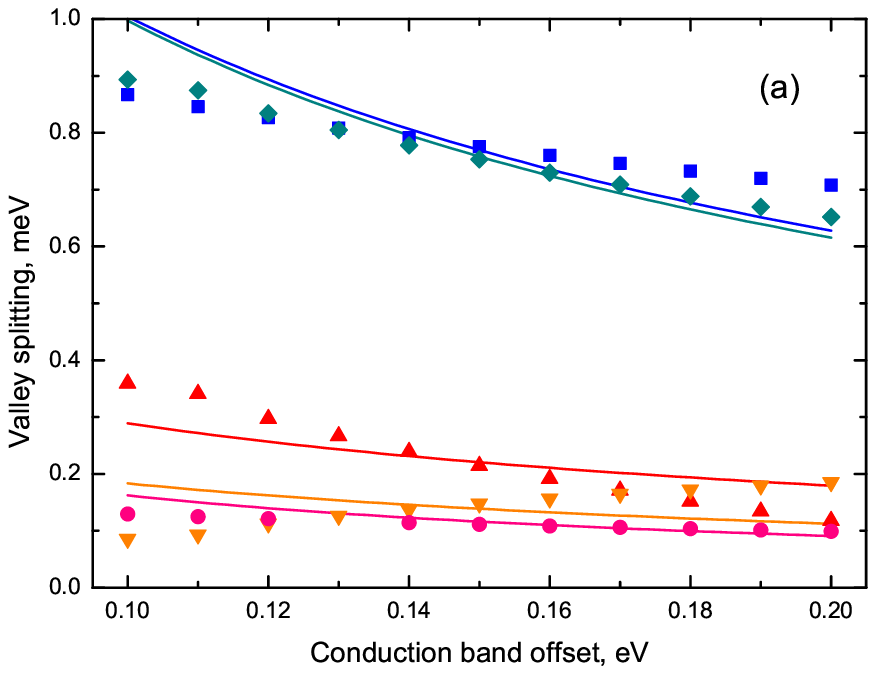}
\hspace{0.5cm}
\includegraphics[width=.29\textwidth]{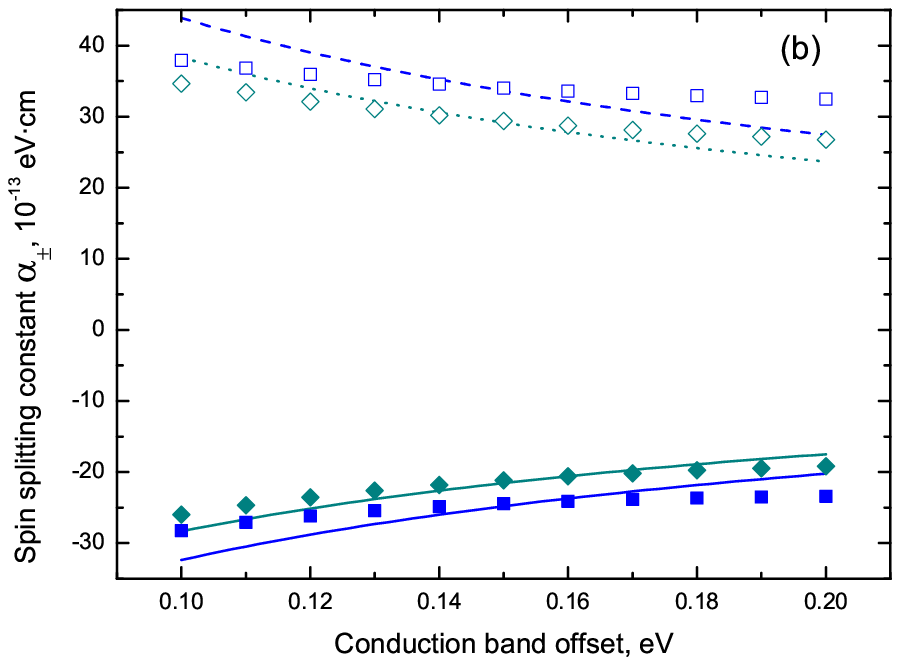}
\caption{Valley (a) and spin (b) splittings for the
lowest conduction subband versus the conduction band offset.
(a) The valley splitting calculated for five QWs with 
N = 31($\blacksquare$), 32($\blacktriangle$), 33($\blacklozenge$), 
34($\blacktriangledown$) and 64({\Large{$\bullet$}}) 
Si monoatomic planes. (b) The spin splitting constants $\alpha_{-}$ 
($\blacksquare$,$\blacklozenge$) and $\alpha_+$ 
($\square$,$\lozenge$) for 
31 ($\blacksquare$,$\square$) and 33($\blacklozenge$,$\lozenge$)
Si atomic layers. 
The fitting by using the extended envelope function approach is shown
by corresponding lines.
}\label{Fig2}
\end{figure}
Figure~2 shows the valley and spin splitting constants as a
function of the conduction band offset for QWs with 31, 32, 33, 34
and 64 Si atomic layers. The fifth structure is taken in order to
provide comparison with Refs.~\onlinecite{Boykin_vo,Friesen}. In
Fig.~2, in addition to the tight-binding calculations, we present
analytical results on valley and spin splitting in the framework
of the extended envelope function approach. A detailed discussion
of the analytical treatment is given in subsection
{\ref{ss:analyt}}. Here we only point out an excellent agreement
between results for the splittings as a function of the QW width
and satisfactory description of the dependence of these splittings
on the band offset.
\subsection{Numerical calculations in the presence of electric field}
Figure~3 demonstrates the variation of spin splitting constants
with the electric field $F_{z}$ for the $e1$ valley-split
subbands. In accordance with symmetry considerations, the
calculations show that the spin splitting becomes anisotropic in
QWs with odd numbers of atomic planes and appears in QWs with even
numbers of atomic planes. The variation of valley splitting is
very weak and we do not present it here.

In order to determine the coefficients $\alpha_{\pm}$ and
$\beta_{\pm}$ we performed the tight-binding calculation of the
spin splittings for the electron wave vectors $\bm{k}\parallel
[110]$ and $\bm{k}\parallel [1\bar10]$ and then applied
Eq.~(\ref{ab}) to find the constants $\alpha$ and $\beta$
directly. The electric field is introduced as a shift of diagonal
energies in the tight-binding Hamiltonian. In accordance with
Eq.~(\ref{TB_field}), we choose electrostatic potential to be a
linear function of $z$ both inside the QW and in the barrier areas
near the interfaces. The area of the constant electric field is
extended inside barriers enough to neglect dependence of the
splitting on the choice of potential profile. Note that the field
values in Fig.~3 are small enough to avoid the tunnelling of an
electron from the QW inside the barrier.

At zero electric field, $\beta$=0 for arbitrary value of $N$ and,
similarly, $\alpha = 0$ for even $N$. In this case the spin
splitting $\Delta_{\rm so}(\bm{k})$ is independent of the
azimuthal angle of the ${\bm k}$ vector. However, with increasing
the field the diversity in values of $\alpha$ and $\beta$ for QWs
with $N=31,32,33,34$ decreases.

The further discussion of spin splitting as a function of electric 
field is continued in the next section. It suffices to note here
that within the chosen range of field values, the coefficients
$\alpha_{\pm}$ in QWs with odd $N$ are linear functions of
$F_{z}^2$. In contrast, in QWs with even numbers of atomic planes
where the spin splitting is absent at zero field, $\alpha_{\pm}$
are proportional to $F_{z}$. We stress that the field-induced
change of $\alpha_{\pm}$ (odd $N$) becomes comparable to the
zero-field value of $\alpha_{\pm}$ in quite weak fields $F_{z}\sim
4 \cdot 10^4$ V/cm.

In the previous paper\cite{Our_SiGe} we developed the extended
envelope function model in order to demonstrate that the spin
splitting in macroscopically symmetrical QWs is fully defined by
interfaces. At non-zero electric field two mechanisms are
possible, namely, the SIA and IIA mechanisms. One of the goals of
current research is to establish the most important term in
realistic QWs. To reveal carefully the comparative role of two
mechanisms we present analytical treatment of results shown in
Figs.~2 and 3 in the framework of the envelope function approach.

\subsection{Extended envelope function approach}\label{ss:analyt}
Here we propose an extended envelope function
approach~\cite{Our_SiGe} to describe valley and spin splittings in
the presence of an external or built-in electric field. The
electron wave function is written as
\begin{equation} \label{psi}
\Psi({\bm r}) =  {\rm e}^{{\rm i} {\bm k}_{\parallel} \cdot {\bm
\rho}} [ \varphi_1(z) \psi_{ {\bm k}_{0} }({\bm r}) + \varphi_2(z)
\psi_{- {\bm k}_{0} }({\bm r})]\:,
\end{equation}
where $\psi_{\pm{\bm k}_{0}}({\bm r}) = {\rm e}^{ \pm{\rm i} k_{0}
z} u_{\pm{\bm k}_{0}}({\bm r})$ is the Bloch function at the the
extremum points $\pm {\bm k}_0$ on the line $\Delta$ in the
Brillouin zone. The spinor envelopes $\varphi_1, \varphi_2$ in
Eq.~(\ref{psi}) are conveniently presented as a four-component
bi\-spin\-or
\begin{equation}
\hat{\varphi}(z) = \left[ \begin{array}{c} \varphi_1(z) \\
\varphi_2(z) \end{array} \right]\:.
\end{equation}
The effective Hamiltonian acting on $\hat{\varphi}(z)$ is written
as a 4$\times$4 matrix consisting of the standard
zero-appr\-ox\-im\-ation Hamiltonian
\begin{equation} \label{h0}
{\mathcal H}_0 = \frac{\hbar^2}{2} \left[ -  \frac{d}{dz}
\frac{1}{m_l(z)} \frac{d}{dz} + \frac{k_x^2 + k_y^2}{m_t(z)}+U(z)
\right]
\end{equation}
which is independent of valley and spin indices, and an
interface-induced $\delta$-functional perturbation
\begin{equation} \label{h1}
{\mathcal H}' = \hat{V}_L \delta(z - z_L) + \hat{V}_R \delta(z -
z_R)\:.
\end{equation}
Here $m_l$ and $m_t$ are the longitudinal and transverse electron
effective mass in the $\Delta$ valley of the bulk material, $z_L$
and $z_R$ are the coordinates of the left- and right-hand-side
interfaces, the potential energy $U(z)$ is referred to the bottom
of the conduction band in Si and given by
\[
U(z) = V \theta_b (z) - e F_z z
\]
with $V$ being the conduction-band offset, $\theta_b(z) = 1$ in
the SiGe barrier layers and $\theta_b(z) = 0$ inside the Si layer.
The explicit form of the matrices $\hat{V}_{L,R}$ obtained by
using symmetry considerations is presented in
Ref.~\onlinecite{Our_SiGe}.

In the zero approximation, i.e., neglecting the valley-orbit and
spin-orbit coupling, $\hat{V}_{L,R} = 0$, the bispinor is given by
\[
\hat{\varphi}(z) = \left[ \begin{array}{c} c_1 \\
c_2 \end{array} \right] \chi(z)\:,
\]
where $c_1, c_2$ are arbitrary $z$-independent spinors and the
function $\chi(z)$ satisfies the Schr{\"o}dinger equation
\begin{equation}\label{h0_sol}
\mathcal{H}_0 \chi(z) = E \chi(z)\:.
\end{equation}
In the following we take into consideration only the lowest
size-quantized electronic subband $e1$.

The next step is an allowance for the interface-induced
spin-independent mixing between the valleys ${\bm k}_0$ and $-
{\bm k}_0$ described by the matrices
\[
\hat{V}_R = \left[ \begin{array}{cc} 0 & \Lambda_{\rm R} \\
\Lambda^*_{\rm R} & 0 \end{array} \right]\:,\:\hat{V}_L =
\left[ \begin{array}{cc} 0 & \Lambda_{\rm L} \\
\Lambda^*_{\rm L} & 0 \end{array} \right]\:,
\]
where $\Lambda_{\rm R} = \lambda {\rm e}^{- {\rm i} k_0 a}$,
$\Lambda_{\rm L} = \lambda {\rm e}^{{\rm i} k_0 a}$, $a = z_R -
z_L = N a_0/4$ is the QW width, $a_0$ is the microscopic lattice
constant, and $\lambda$ is a complex coefficient.

In terms of the envelope $\chi(z)$ the matrix element of valley
mixing can be written as
\begin{multline}\label{v_splitting}
\langle 1 \vert \mathcal{H}' \vert 2 \rangle = \left| \lambda
\right| \left[ ( \chi_L^2 + \chi_R^2 ) \cos(k_0 a -
\phi_{\lambda}) \right. \\ 
\left.+ \right. \left. {\rm i}( \chi_L^2-\chi_R^2 )\sin(k_0 a -
\phi_{\lambda}) \right]\:,
\end{multline}
where $|\lambda|$ and $\phi_{\lambda}$ are the modulus and the
phase of $\lambda$, $\chi_{R, L}$ are the values of the envelope
$\chi$ at the right and left interfaces, $\chi(\pm a/2)$,
respectively. The valley-orbit split states have the energy
$E_{\pm} = E_0 \pm |\langle 1 \vert \mathcal{H}' \vert 2
\rangle|$, where $E_0$ is the eigen energy of Eq.~(\ref{h0_sol}),
so that the splitting is equal to
\begin{multline} \label{vo}
\Delta_{ {\rm v}} = 2 |\langle 1 \vert \mathcal{H}'
\vert 2 \rangle| \\= 2 |\lambda| \sqrt{\chi^4_L + \chi^4_R + 2
\chi^2_L \chi^2_R \cos{[2(k_0 a - \phi_{\lambda})]}}\:.
\end{multline}
The bispinors for the upper ($+$) and lower ($-$) states are given
by
\begin{equation} \label{varphivo}
\hat{\varphi}_{\pm, s}(z) = \frac{1}{\sqrt{2}} \left[ \begin{array}{c} c_s \\
\pm {\rm e}^{- {\rm i} \phi_M}c_s \end{array} \right]\:,
\end{equation}
where $c_s$ is spinor $\uparrow$ for the electron spin $s= 1/2$
and $\downarrow$ for the electron spin $s= -1/2$, $\phi_M$ is the
phase of the matrix element (\ref{v_splitting}).

The tight-binding calculations show that the spin splitting is
much smaller as compared to the valley splitting $\Delta_{ {\rm
v}}$. Therefore, the spin splitting can be
considered independently for the upper and lower valley-orbit
split states (\ref{varphivo}). The corresponding matrix elements
are reduced to
\begin{multline}\label{hso}
{\cal H}'_{ss'}(\bm{k};e1,\pm) = M_{1s,1s'} \\
\pm \Bigl(\rm{Re}\{M_{1s, 2s' }\} \cos{\phi_M}  
+ \rm{Im}\{M_{1s,2s'}\}\sin{\phi_M}\Bigr),
\end{multline}
where the subscript indices $1,2$ enumerate the valleys ${\bm
k}_0$, $- {\bm k}_0$ and $s,s' = \pm 1/2$ are the spin indices,
the components $M_{1s,1s'}, M_{1s, 2s'}$ written as 2$\times$2
matrices $M_{11}, M_{12}$ are related to similar 2$\times$2
matrices $V_{R,11}$, $V_{L,11}$, $V_{R,12}$, $V_{L,12}$ by
\[
M_{11} = \chi_L^2 V_{L,11} + \chi_R^2 V_{R,11},\;\;\; M_{12} =
\chi_L^2 V_{L,12}+\chi_R^2 V_{R,12}\ .
\]
The spin-dependent contributions to $V_{R,11}$, $V_{L,11}$, $V_{R,12}$,
$V_{L,12}$ are linear combinations of $h({\bm k})$ and $h'({\bm
k})$, see Ref.~\onlinecite{Our_SiGe}:
\begin{widetext}
\begin{equation} \label{VR}
\hat{V}_R = \left[ \begin{array}{cc} S~ h({\bm k}) + S'~ h'({\bm
k}) & [\lambda  + p~ h({\bm k}) + p'~ h'({\bm k})] {\rm e}^{- {\rm
i} k_0 a}
\\
\left[ \lambda^* + p^*~ h({\bm k}) + p^{\prime *}~ h'({\bm k})
\right] {\rm e}^{{\rm i} k_0 a} & S~ h({\bm k}) + S'~ h'({\bm k})
\end{array} \right]
\end{equation}
\end{widetext}
and similar equation for $\hat{V}_L$ with the coefficients
interrelated with $\lambda, S, S', p, p'$ due to the
mirror-rotation operation ${\cal S}_4$ (odd $N$) or the inversion
$i$ (even $N$) which transforms the right-hand-side interface into
the left-hand-side one. Taking into account the relation between
coefficients entering the matrices $\hat{V}_R$ and $\hat{V}_L$ and
the notations of Ref.~\onlinecite{Our_SiGe} we can write the spin
Hamiltonians (\ref{hso}) in the form of Eq.~(\ref{hamso}), namely,
\begin{equation} \label{hsopm}
\mathcal{H}^{(1)}(\bm{k};e1,\pm) = \alpha_{\pm} h(\bm{k}) + 
\beta_{\pm} h'(\bm{k})\:.
\end{equation}
For the coefficients $\alpha_{\pm}, \beta_{\pm}$ describing the
spin splitting of the valley-orbit split subbands, we obtain
\begin{equation} \label{ss}
\alpha_{\pm} = [\chi^2_R - (-1)^N \chi^2_L] S \pm |p| H_{\alpha}
(\phi_p)\:,
\end{equation}
\[
\beta_{\pm} = (\chi^2_R - \chi^2_L) S' \mp |p'|
H_{\beta}(\phi_{p'})\:.
\]
Here
\[
H_{\alpha}(\phi) = \chi^2_R \cos{(k_0 a - \phi + \Phi)} - (-1)^N
\chi^2_L \cos{(k_0 a - \phi - \Phi)}
\]
\[
H_{\beta}(\phi) = \chi^2_R \cos{(k_0 a - \phi + \Phi)} - \chi^2_L
\cos{(k_0 a - \phi - \Phi)}\:,
\]
\[
\Phi = {\rm arg} \left( \chi^2_L {\rm e}^{{\rm i} (k_0 a -
\phi_{\lambda})} + \chi^2_R {\rm e}^{-{\rm i} (k_0 a -
\phi_{\lambda})} \right) \:,
\]
the parameters $S, S'$ describe the intra-valley contributions to
the interface-induced electron spin mixing, and the parameters $p
= |p|{\rm e}^{\phi_p}$, $p'=|p'| {\rm e}^{\phi_{p'}}$ describe the
spin-dependent inter-valley mixing. Oscillatory dependence of the
valley and spin splittings on the QW thickness $a$ is caused by
interference of electron waves arising from inter-valley
reflection off the left- and right-hand side interfaces.

\subsection{Comparison between tight-binding and envelope function
approach}
In the absence of an electric field, one has $\chi^2_L =
\chi^2_R$, $\Phi = 0$ for positive and $\Phi = \pi$ for negative
values of $\cos{(k_0 a - \phi_{\lambda})}$, and Eqs.~(\ref{vo}),
(\ref{ss}) reduce to\cite{Our_SiGe}
\[
\Delta_{{\rm v}} = 4 \chi^2_L |\lambda\ \cos{(k_0 a
- \phi_{\lambda})}|\:,
\]
$\alpha_{\pm} = \beta_{\pm} = 0$ for even $N$, and
\[
\alpha_{\pm} = 2 \chi^2_L [ S \pm \eta |p| \cos{(k_0 a -
\phi_p)}]\:,\: \beta_{\pm} = 0
\]
for odd $N$, where $\eta = {\rm e}^{{\rm i} \Phi} = {\rm
sign}\{\cos{(k_0 a - \phi_{\lambda})}\}$. The curves in Fig.~1 are
calculated by using the following best-fit set of parameters:
$|\lambda| = 65$ meV$\cdot$\AA, $\phi_{\lambda} = $0.013$\pi$,
$|p|$ =$4450$ meV$\cdot$\AA$^2$, $\phi_p = 0.095 \pi$, 
$S$ = 650 meV$\cdot$\AA$^2$.

Although tight-binding-model values of the coefficients in the
present work are quite different from the previous estimations,
the envelope function approach proves its adequate description of
the valley and spin splittings as a function of the QW width. The
analytical approach with merely five parameters perfectly fits the
complex microscopical calculation.

Comparison of new results with experimental data of Wilamovsky et
al.\cite{Janch} shows the better agreement. With the necessary
correction\cite{Glazov} the experimental results give
$\sim$$0.34\times 10^{-12}$ eV$\cdot${\AA}~ for the spin splitting
constant in a 120{\AA}-thick QW. The more detailed comparison
should be done with caution since effects of disorder and built-in
electric fields can have crucial influence. However, the
coincidence in an order of magnitude shows that our calculations
agree with the available experimental data.

According to Fig.~2, the description of dependence of spin
splitting on the band offset is not so perfect in the framework of
envelope function approach. In fact, the variation in band offset
results in a complex (obviously non-linear) behaviour of the
parameters in the boundary conditions at the interfaces.

Figure~3 is the main result of this work and we discuss it in more 
details. The agreement between the two approaches seen in Fig.~3a
shows that the extended envelope function approach catches the
physics of spin splitting induced by the applied electric field.
Moreover, the fact that this agreement takes place with no
addition of a SIA term shows that in the system under study the
IIA contribution is dominating. It should be noted that Eq.
(\ref{ss}) for $\alpha_{\pm}$ contains only parameters which can
be extracted from Fig.~1, and, indeed, Fig.~3a does not contain
any fitting parameters. Additional fitting parameters used to
describe Fig.~3b are as follows: $|p'|$ =$700$ meV$\cdot$\AA$^2$,
$\phi_{p'} =  \pi$ and $S'$ = $70$ meV$\cdot$\AA$^2$.

In the high-field limit $\chi_L \chi_R/(\chi^2_L + \chi^2_R) \to
0$ so that either $\chi^2_L \ll \chi^2_R$ or $\chi^2_L \gg
\chi^2_R$, and Eqs.~(\ref{vo}), (\ref{ss}) transfer to
\[
\Delta_{ {\rm v}} = 2 |\lambda|\ {\rm
max}\{\chi^2_L, \chi^2_R\}\:,
\]
\[
\alpha_{\pm} = 2\ {\rm sign}\{F_z^{N+1}\} [ S \pm |p| \cos{(\phi_p
- \phi_{\lambda})}]\ {\rm max}\{\chi^2_L, \chi^2_R\} \:,
\]
\[
\beta_{\pm} = 2\ {\rm sign}\{F_z\}\ [ -S' \mp |p'| \cos{(\phi_{p'}
- \phi_{\lambda})}]\ {\rm max}\{\chi^2_L, \chi^2_R\}\:.
\]
Since one of the interfaces becomes inaccessible to the electron
the oscillatory behavior vanishes in strong fields.

It should be stressed that the parity of the coefficients
$\alpha_{\pm}$ and $\beta_{\pm}$ following from the above
equations completely agrees with the general symmetry
considerations, see Eqs.~(\ref{evenodd}) and (\ref{D2h_field_Hso}). We
also note that a monoatomic shift as a whole of the QW position in
the structure results in an inversion of sign of $\alpha_{\pm}$
while the values of $\beta_{\pm}$ remain unchanged.

\begin{figure}
\includegraphics[width=.335\textwidth]{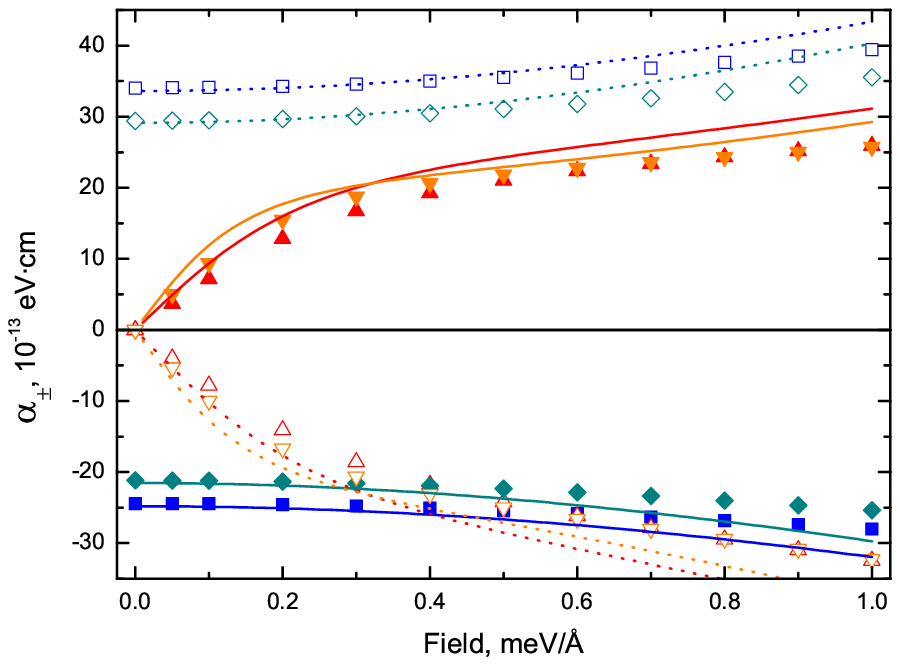}
\hspace{0.5cm}
\includegraphics[width=.33\textwidth]{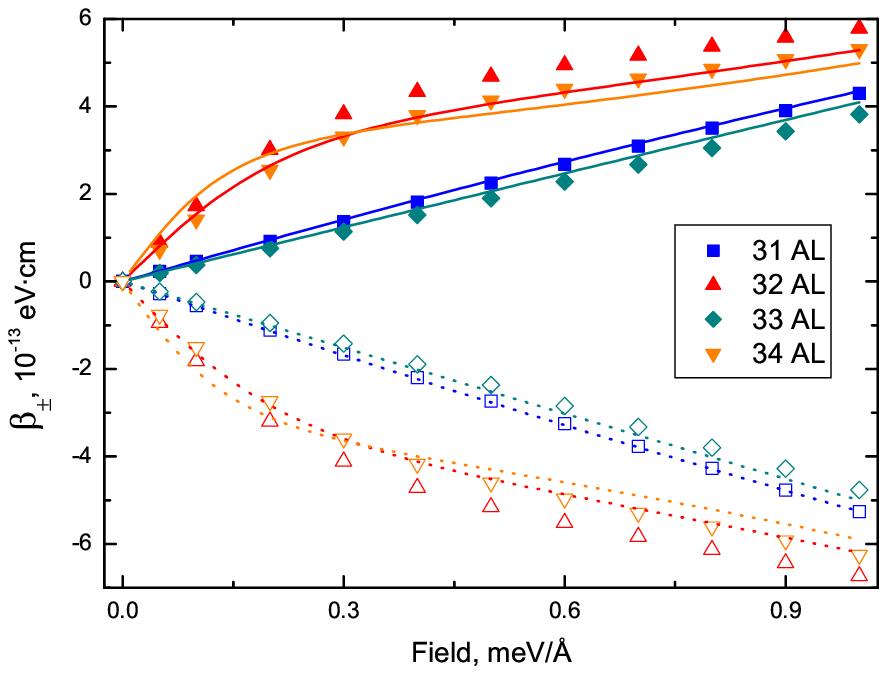}
\caption{Spin-splitting constants $\alpha_{\pm}$ and $\beta_{\pm}$ 
for the lowest conduction subband versus the electric field $F_z$
calculated for four QWs with 31, 32, 33 and 34 Si monoatomic
planes. Points are calculated in the $sp^3d^5s^*$ tight-binding model. Lines
represent fitting by using the extended envelope function approach.}
\end{figure}
\section{Conclusion}

The $sp^3d^5s^*$ tight-binding model has been used to calculate
the electron dispersion in heterostructures grown from multivalley
semiconductors with the diamond lattice, particularly, in the
Si/SiGe structures. The model allows one to estimate
quantitatively the valley and spin splittings of electron states
in the quantum-confined ground subband as well as the
electric-field dependence of the spin splitting. In the employed
tight-binding model, this splitting is mostly determined by the
spin-dependent mixing at the interfaces. As a result the
coefficients describing the Dresselhaus term in unbiased QWs are
oscillating functions of the odd number $N$ of Si monoatomic
layers. Under an electric field applied along the growth axis a
non-zero Rashba term appears in QWs with both even and odd Si
atomic layers. In small fields, the Dresselhaus term is linear in
the structures with even $N$ ($D_{2h}$ point group) and quadratic
in structures with odd $N$ ($D_{2d}$ point group). Thus, in quite
low fields about $10^{-4}$ eV$\cdot$cm the spin splitting becomes
anisotropic and oscillations as a function of the QW width are
suppressed.
In addition to numerical calculations, an extended envelope
function approach is utilized to interpret the results of
tight-binding calculations. The inclusion of spin-dependent
reflection of an electronic wave at the interface and
interface-induced inter-valley mixing permits one to describe
quite well the numerical dependencies of the valley-orbit and
spin-orbit splittings upon the number of Si atomic planes and the
electric field.

\acknowledgments{This work was financially supported by
RFBR and CNRS PICS projects, programmes of RAS and ``Dynasty'' 
Foundation --- ICFPM.}

\newpage

\bibliography{bibliography}

\end{document}